\documentclass[twocolumn,final,twoside,conference,letterpaper]{IEEEtran}

\usepackage[nolist]{acronym}
\usepackage{algorithm}
\usepackage{algpseudocode}
\usepackage{amsfonts}
\usepackage{amsmath}
\usepackage{amssymb}
\usepackage{amsthm}
\usepackage{bm}
\usepackage{booktabs}
\usepackage{cases}
\usepackage{cite}
\usepackage{color}
\usepackage[dvipsnames]{xcolor}
\usepackage{csquotes}
\usepackage{dsfont}
\usepackage{epsfig}
\usepackage{graphicx}
\usepackage{hhline}
\usepackage{lipsum}
\usepackage{mathrsfs}  
\usepackage{mathtools}
\usepackage{mathtools}
\usepackage{multirow}
\usepackage{perso}
\usepackage{pgfplots}
\usepackage{psfrag}
\usepackage{tikz}
\usepackage[normalem]{ulem}
\usepackage{xspace}
\usepackage{soul}

\usetikzlibrary{plotmarks}
\usetikzlibrary{spy}
\usetikzlibrary{chains}
\usetikzlibrary{arrows.meta,
                positioning,
                quotes
                }

 \pgfplotsset{compat=newest}
    \pgfplotsset{plot coordinates/math parser=false}
    \pgfplotsset{
    label style={anchor=near ticklabel},
    xlabel style={yshift=0.0em},
    ylabel style={yshift=-0.3em},
    tick label style={font=\footnotesize },
    label style={font=\footnotesize},
    legend style={font=\footnotesize},
    title style={font=\fontsize{7}}}

\mathtoolsset{showonlyrefs}

\usepackage{placeins}
\usepackage{mathtools}

\usepackage{etoolbox}

\usepackage{flushend}

\usetikzlibrary{decorations.fractals}

\usepackage{academicons}
\definecolor{orcidlogocol}{HTML}{A6CE39}
\definecolor{forestgreen}{rgb}{0.13, 0.55, 0.13}
\usepackage{orcidlink}
\hypersetup{
  colorlinks=false,
  linkbordercolor=white,
 urlbordercolor=white,
pdfborder={0 0 0}
}

\newcommand{\rank}{\mathrm{rank}}

\DeclareMathOperator*{\argmax}{arg\,max}

\newcommand{\T}{{\intercal}}
\newcommand{\cmmnt}[1]{}

\newcommand{\CA}{$\mathcal{C}_{\mathrm{rep\,}}^{(21,7)}$}
\newcommand{\CB}{$\mathcal{C}_{\mathrm{red\,}}^{(21,7)}$}
\newcommand{\CC}{$\mathcal{C}_{\mathrm{con\,}}^{(24,7)}$}
\newcommand{\CD}{$\mathcal{C}_{\mathrm{red\,}}^{(24,7)}$}
\newcommand{\CE}{$\mathcal{C}_{\mathrm{rep\,}}^{(28,7)}$}
\newcommand{\CF}{$\mathcal{C}_{\mathrm{con\,}}^{(28,7)}$}

\newcommand{\CG}{$\mathcal{C}_{\mathrm{t-red\,}}^{(33,8)}$}

\newcommand{\CH}{$\mathcal{C}_{\mathrm{rep\,}}^{(24,8)}$}
\newcommand{\CI}{$\mathcal{C}_{\mathrm{red\,}}^{(24,8)}$}
\newcommand{\CJ}{$\mathcal{C}_{\mathrm{con\,}}^{(27,8)}$}
\newcommand{\CK}{$\mathcal{C}_{\mathrm{red\,}}^{(27,8)}$}
\newcommand{\CL}{$\mathcal{C}_{\mathrm{rep\,}}^{(32,8)}$}
\newcommand{\CM}{$\mathcal{C}_{\mathrm{red\,}}^{(32,8)}$}

\IEEEoverridecommandlockouts 
\begin{document}

	\begin{acronym}
\acro{BSC}{binary symmetric channel}
\acro{LDPC}{low-density parity-check}
\acro{MAP}{maximum a posteriori}
\acro{deg-MAP}{degenerate \ac{MAP}}
\acro{PCM}{parity-check matrix}
\acro{CSS}{Calderbank-Shor-Steane}
\end{acronym}

	\title{Optimal Single-Shot Decoding of Quantum Codes}

\author{
    \IEEEauthorblockN{Aldo Cumitini\orcidlink{0009-0006-5962-4880}, Stefano Tinelli\orcidlink{0009-0008-2336-5885}, Bal\'azs Matuz\orcidlink{0000-0002-0133-6564}, Francisco L\'azaro\orcidlink{0000-0003-0761-7700}, Luca Barletta\orcidlink{0000-0003-4052-2092}}
    \vspace{-0.6cm}
    
\thanks{Stefano Tinelli, Bal\'azs Matuz and Francisco L\'azaro are with the Institute of Communications and Navigation of DLR (German Aerospace Center),    Wessling, Germany. (email: \{stefano.tinelli, balazs.matuz, francisco.lazaroblasco\}@dlr.de) Corresponding author: Stefano Tinelli.}
\thanks{Aldo Cumitini and Luca Barletta are with Dipartimento di Elettronica, Informazione e Bioingegneria, Politecnico di Milano, Milano, Italy. (email: aldo.cumitini@mail.polimi.it, luca.barletta@polimi.it)}
\thanks{The two first authors contributed equally and are listed in alphabetical order.}
\thanks{Copyright \copyright 2024 IEEE. Personal use of this material is permitted.
However, permission to use this material for any other purposes must be
obtained from the IEEE by sending a request to pubs-permissions@ieee.org}
    }

\thispagestyle{empty} \pagestyle{empty}

	\maketitle

\begin{abstract}		
 We discuss single-shot decoding of quantum \acl{CSS} codes with faulty syndrome measurements. We state the problem as a joint source-channel coding problem. By adding redundant rows to the code's parity-check matrix, we obtain an additional syndrome error correcting code which addresses faulty syndrome measurements. Thereby, the redundant rows are chosen to obtain good syndrome error correction capabilities while keeping the stabilizer weights low. Optimal joint decoding rules are derived which, though too complex for general codes, can be evaluated for short quantum codes.
	\end{abstract}
 \vspace{3pt}

\section{Introduction}

Recently, quantum information technologies have attracted great interest, since for certain applications they promise significant advantages compared to conventional technologies. One prominent example is Shor's algorithm for finding the prime factors of an integer \cite{Shor07}, which provides an exponential speed-up compared to the best known classical algorithm.
A major challenge for quantum computers is decoherence, i.e., the unintended interaction of qubits with their environment that leads to a loss of quantum information. This calls for powerful quantum error correction schemes. Although the noisy codeword cannot be observed directly, syndrome measurements can be performed using ancilla qubits to extract information about errors that affect a quantum system \cite{lidar2013quantum}. However, the quantum circuits used to extract these syndrome measurements are themselves faulty. Thus, one has to deal with both qubit and syndrome measurement errors. 

A straightforward approach to combat syndrome errors is to repeat the syndrome measurements multiple times \cite{shor1996fault}, a process known as Shor's syndrome extraction. In Shor's syndrome extraction, the number of measurement repetitions has to scale linearly with the code distance in order to achieve fault tolerance. An alternative is to rely on so-called single shot error correction \cite{bombin2015single}, which implies carrying out redundant syndrome measurements, which are not necessarily repetitions of previously carried out measurements, but rather linear combinations thereof \cite{Ashikhmin}. 

Such linear combinations might be subject to higher measurement uncertainty, but when employing them, it is sometimes possible to achieve fault tolerance using only a constant number of measurement rounds \cite{bombin2015single}.

This work focuses on single-shot decoding of quantum error-correcting codes. In particular, by stating the problem as a joint source-channel coding problem, we gain further insight into the construction of the syndrome error-correcting code. We derive the optimal joint decoding rule (for the qubit and syndrome codes) as well as a relaxation thereof that ignores error degeneracy. The evaluation of the resulting expressions is, in general, complex, albeit feasible for small codes. Finally, the experimental results illustrate the performance of different syndrome error-correcting code constructions. 

\section{Quantum Error Correction}
We consider $[[n_q,k_q]]$ \ac{CSS} codes \cite{Calderbank_1996}. The code constraints can be represented by a binary   $ (n_q-k_q) \times 2n_q$ parity-check matrix of form
\begin{equation} \label{eq:CSS}
    \bm{H}_q = \begin{bmatrix}
        \bm{H}_X & \bm{0}\\
        \bm{0} & \bm{H}_Z
    \end{bmatrix} .
\end{equation}
The  $(n_q - k_x)\times n_q$ and $(n_q - k_z) \times n_q$ sub-matrices $\bm{H}_X$ and $\bm{H}_Z$ (with $k_q = k_x + k_z - n_q$) 
must fulfill $\bm{H}_X\bm{H}_Z^\T = \bm{0}$ to comply with the commutation requirement of the stabilizers.

In quantum systems, it is not possible to measure the qubits without perturbing the state. Instead, quantum error correction is performed relying on so-called (quantum) syndrome measurements that yield a syndrome vector $\bm{s}_q$. This vector can be expressed as
\begin{equation}
\label{eq:SymProd}
\bm{s}_q^\T = \bm{H}_q \bm {e}_q^\T
\end{equation}
where $\bm e_q=[\bm e_Z|\bm e_X]$ is a binary vector of length $2n_q$ uniquely associated with a Pauli error. 
The $X$ and $Z$ components of the error vector are swapped in \eqref{eq:SymProd} in order to implement the symplectic product \cite{MacKay_2004}.
In particular, when the $i$-th qubit is subject to a Pauli $X$ error, the $i$-th element in $\bm {e}_X$ is set to one, whereas when it is subject to a Pauli $Z$ error, the $i$-th element of $\bm {e}_Z$ is set to one. 

\section{System Model}
As in \cite{MacKay_2004}, we model the channel error vector $\bm e_q=[\bm e_Z|\bm e_X]$ as the output  of a \ac{BSC} which introduces independent $Z$ and $X$ errors with the same probability $\epsilon$. Due to the independence of $X$ and $Z$ errors, we can decode them independently using the matrices $\bm{H}_Z$ and $\bm{H}_X$, respectively. To simplify notation, in the following, we drop the subscripts $X$ and $Z$. Thus, we denote by $\bm{H}$ the $(n-k)\times n$  binary parity-check matrix of an $(n,k)$ linear block code $\mathcal C$. The error vector is denoted by $\bm{e}$. The (error-free) syndrome $\bm s$ is computed as $\bm s=\bm e \bm{H}^\T$.

\begin{figure*}[t]
    \centering
  \includegraphics[width=\textwidth]{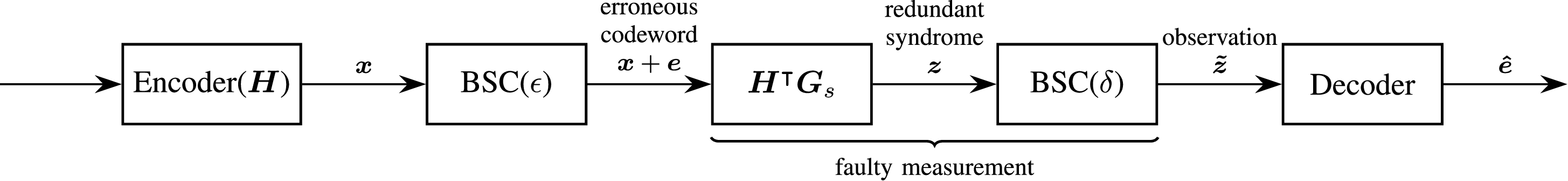}\vspace{-2mm}
    \caption{
    Classical error model. Pauli $X$ and $Z$ errors are assumed to be independent events, resulting in an associated binary error vector $\bm e_q=[\bm e_Z|\bm e_X]$. It can be seen as the output of a \ac{BSC} with error probability $\epsilon$, where $\bm e_Z$ and $\bm e_X$ can be inferred independently. Therefore, the figure only shows one component where the subscripts $X$ and $Z$ are omitted. The vector $\bm e$ is never actually observed. Instead, a corrupted version $\tilde{\bm{z}}$ of the encoded syndrome $\bm{z}=\bm{e} \bm{H}^\T \bm{G}_{s}$ is measured. Measurement errors are modeled by a \ac{BSC} with error probability $\delta$. The decoder provides an estimate $\hat{\bm{e}}$ of the error vector $\bm e$.}
    \label{fig:final_model} \vspace{-2.5mm}
\end{figure*}

In quantum systems, not only the qubits are subject to errors, but also the syndrome measurements can be faulty. In this work, we model errors in syndrome measurements as the transmission of the syndrome over a \ac{BSC} with error probability $\delta$. To provide resilience against syndrome errors, $\bm{s}$ is encoded using an $(m,n-k)$ binary linear code $\mathcal C_s$ with an $(n-k)\times m$ generator matrix $\bm{G}_{s}$.  This yields a redundant (or encoded) syndrome $\bm{z}$. Figure~\ref{fig:final_model} illustrates the abstracted model of our transmission system. 

\subsection{Syndrome Error Probability} \label{sec:syndrome_error_model}
The rows of matrix $\bm{H}$ are denoted as $\bm{h}_1,\ldots,\bm{h}_{n-k}$. In the field of quantum error correction, these rows are also known as stabilizers of the code. Each stabilizer is connected to a syndrome measurement. 
In order to perform the syndrome measurement associated with the $i$-th stabilizer $\bm{h}_i$, typically an ancilla-qubit is injected, and it needs to interact with $\mathrm{w}(\bm{h}_i)$ data qubits, where $\mathrm{w}(\bm{h}_i)$ denotes the Hamming weight of $\bm{h}_i$. 
{A simple error model for syndrome measurement errors is obtained assuming that each of these interactions fails with a given probability $q$ \cite{Ashikhmin}, yielding the following syndrome measurement error probability}
\begin{align} 
\Pr(z_j \neq \tilde{z}_j)
&=\sum_{i \mbox{ \small{is odd}}} \binom{\mathrm{w}(\bm{h}_j)}{i}
q^i (1-q)^{\mathrm{w}(\bm{h}_j)-i}.
\label{eq:sj_error}
\end{align}

Observe that the probability in~\eqref{eq:sj_error} increases with the Hamming weight of $\bm{h}_j$.

{Since $\bm{h}_j$ for $j\in \{1,\ldots, m\}$ may not have constant weights, it is convenient to define the  average error probability $\delta$ as }
\begin{align}\label{eq:avg_delta}
\delta=\frac{\sum_{j=1}^{m} \Pr(z_j \neq \tilde{z}_j)}{m}
\end{align}
{which is the syndrome error probability assumed throughout this work.}

\section{Syndrome Error Correcting Code}

Elaborating on the redundant syndrome $\bm{z}$, we obtain
\begin{align}
\bm{z} &=\bm{s} \bm{G}_{s} = \bm{e} \bm{H}^\T \bm{G}_{s} = \bm{e} \bm{H}^\T_{\text{o}} 
\label{eq:joint_src_channel}
\end{align}
By definition, we have $\rank(\bm{G}_s)=n-k$ and $\rank(\bm{H})=n-k$. 
Exploiting well-known results from linear algebra, it follows that 
matrix $\bm{H}_{\text{o}}= {\bm{G}_{s}}^\T \bm{H}$ has size $m \times n$ and  $\rank(\bm{H}_{\text{o}})=n-k$, since  
\begin{align}
     \rank(\bm{H}_{\text{o}}) & \leq \min(\rank(\bm{G}_s) , \rank(\bm{H})) \\
    \rank(\bm{H}_{\text{o}}) & \geq \rank(\bm{G}_s) + \rank(\bm{H})  - (n - k) ,
\end{align}
where the lower bound on the rank of $\bm{H}_{\text{o}}$ is also known as Sylvester's inequality. 

We make the following observations. First, the $m \times n$ matrix $\bm{H}_{\text{o}}$, $m>n-k$, is  overcomplete, i.e., it contains linearly dependent rows. These linearly dependent rows enable correction of syndrome errors. Second, \eqref{eq:joint_src_channel} describes a joint source-channel coding problem \cite{fresia2010joint}, where $\bm e$ is first compressed with the help of $\bm H$ and then $\bm s$ is encoded to $\bm z$.

\subsection{Code Construction}

Let
\begin{equation} \label{eq:H|P}
    \bm{H}_{\text{o}} = 
 \left[   \begin{array} {c}
        \bm{H} \\ 
        \hline \bm{P}
    \end{array}\right]
\end{equation}
where the $(m-n+k)\times n$ matrix $\bm{P}$ represents the redundant part of $\bm{H}_{\text{o}}$. 
Note that any matrix $\bm{H}_o$ can be rearranged as in~\eqref{eq:H|P}, e.g., by means of Gaussian elimination that identifies $n-k$ linearly independent rows. 
Then, the generator matrix $\bm{G}_s=[\bm I| \bm A]$ (in systematic form) of the syndrome error-correcting code is the solution of
\begin{equation} \label{eq:matA}
    \bm{A}^\T\bm{H} = \bm{P} .
\end{equation}
One may use Gaussian elimination to solve~\eqref{eq:matA} for $\bm A$. 

In the sequel, we assume that the stabilizers of the quantum error-correcting code $\mathcal C$, hence $\bm H$, are given. Given this constraint, our focus lies on the development of the syndrome error correcting code denoted as $\bm{G}_s$.
The classical code design approach is to find a code $\mathcal C_s$ with good distance and thus good error correction properties. However, in the quantum setting, we aim for matrices $\bm{H}_{\text{o}}$ with low-weight rows that not only facilitate implementation, but also minimize the probability of syndrome measurement error $\delta$ (see Section~\ref{sec:syndrome_error_model}). More precisely, we would like to ensure that $\bm P$ is sparse, which is not necessarily guaranteed even when $\bm H$ and $\bm{G}_s$ are sparse.

 In this work, starting from $\bm H$ we generate low-weight redundant rows. The problem of finding a sparse representation of a code can be addressed, e.g., by relying on probabilistic approaches as in \cite{Leon88}. For the short code examples in Section~\ref{sec:results} we can directly exploit the structure of $\bm H$ and construct $m'>m$ low-weight redundant rows (which form $\bm P$) . Alternatively, with a combinatorial approach, it is possible to investigate any possible combination of $j$ rows with $j\in\{2,n-k\}$.
 Once $\bm P$ is obtained, by solving \eqref{eq:matA}, the matrix $\bm A'$ is evaluated. 
 Consequently, the generator matrix $\bm{G'}_s=[\bm I| \bm A']$ of  an $(m', n-k)$ code $\mathcal C_s'$ is constructed. 
  Finally, we select a sub-space of $m$ rows of $\bm P$ to obtain an $(m, n-k)$ code ${\cal C}_s$. For the codes under consideration, the selection of the $m$ rows among the $m'$ candidates can be done by an exhaustive search to maximize the minimum distance (and minimize the multiplicity of minimum weight codewords)  of ${\cal C}_s$.  Alternatively, we will also provide examples of a concatenation of the syndrome error correcting code with a repetition code, since it is always possible to repeat the syndrome measurements.

\section{Decoding}
\subsection{Degenerate Maximum A Posteriori Decoding}
In the quantum setting, 
 two error vectors $\bm{e}$ and $\tilde{\bm{e}}$ are said to be degenerate if their modulo-2 sum $\bm{e}+\tilde{\bm{e}}$ is a stabilizer, that is, if it belongs to the row span of $\bm{H}$. Degenerate errors are indistinguishable from each other and lead to the same quantum state (see  \cite{PouInter} for a detailed discussion on degeneracy).
Thus, it is possible to group the error operators into cosets $\mathcal{E}$, which can be thought of as equivalence classes. All errors in a coset $\mathcal{E}$ can be corrected by the same recovery operator.  The task of a degenerate decoder is to identify the right coset and apply the respective correction to the corrupted state. As in \cite{PouInter}, the decoder aims to find the coset that minimizes the probability of error given the syndrome, extending the optimal decoding rule, e.g. \ac{MAP}, to the quantum scenario.

Hence, a degenerate \ac{MAP} decoder computes the most probable coset given the noisy syndrome observation $\tilde{\bm{z}} $ as
\begin{align}
   \hat{\mathcal{E}} &= \argmax_{\mathcal{E}} \Pr(\mathcal{E}|\Tilde{\bm{z}} )
                    = \argmax_{\mathcal{E}}\frac{\Pr(\Tilde{\bm{z}} |\mathcal{E})\Pr(\mathcal{E})}{\Pr(\Tilde{\bm{z}})} \label{eq:Bayes}\\
                    &= \argmax_{\mathcal{E}}{\Pr(\Tilde{\bm{z}} |\mathcal{E})\Pr(\mathcal{E})}. \label{eq:prop}
\end{align}
Elaborating on $\Pr(\Tilde{\bm{z}} |\mathcal{E})$ we obtain
\begin{align}
    \Pr(\Tilde{\bm{z}}| \mathcal{E}) &=  \frac{\Pr(\Tilde{\bm{z}}, \mathcal{E})}{\Pr(\mathcal{E})}
   \stackrel{(a)}{=} \frac{1}{\Pr(\mathcal{E})}\sum_{\bm{e}\in\mathcal{E}}\Pr(\Tilde{\bm{z}},\bm{e}) 
   \\ 
    & = \frac{1}{\Pr(\mathcal{E})}\sum_{\bm{e}\in\mathcal{E}}\Pr(\Tilde{\bm{z}}|\bm{e})\Pr(\bm{e}) \label{eq:Bayes3},
\end{align}
where in $(a)$ we exploited the fact that the error events in $\mathcal E$ are all disjoint. 
Inserting~\eqref{eq:Bayes3} into~\eqref{eq:prop} we obtain

\begin{equation}
    \hat{\mathcal{E}} = \argmax_{\mathcal{E}}{\sum_{\bm{e}\in\mathcal{E}}\Pr(\Tilde{\bm{z}}|\bm{e})\Pr(\bm{e})}. \label{eq:degMAP}
\end{equation}
Note that evaluating the expression in~\eqref{eq:degMAP} requires  processing of all $2^n$ different error vectors, and is thus only feasible for small values of $n$.

In this paper, the following error model is assumed.
Let $\mathrm  d(\bm{z},\Tilde{\bm{z}})$ be the Hamming distance between the vectors $\bm{z}$ and $\Tilde{\bm{z}}$. 
The probability associated with an error vector $\bm e$ over a \ac{BSC} with crossover probability $\epsilon$ is defined as,
\begin{equation}
  \Pr(\bm e)=\left( \frac{\varepsilon}{1-\varepsilon} \right)^{\mathrm{w}(\bm{e})} (1-\varepsilon)^n .\label{eq:pe}
\end{equation}
Similarly, over a \ac{BSC} with crossover probability $\delta$, we have
\begin{equation}
  \Pr(\Tilde{\bm{z}}|\bm{e}) = \Pr(\Tilde{\bm{z}}|\bm{z}(\bm e)) = \left( \frac{\delta}{1-\delta} \right)^{\mathrm d(\bm{z}(\bm e),\Tilde{\bm{z}})} (1-\delta)^{m} \label{eq:pze},
\end{equation}
where $\bm z(\bm e)$ is the redundant syndrome vector induced by the error pattern $\bm e$.

\subsection{Maximum A Posteriori Decoding}
Ignoring the effect of degeneracy, a classical \ac{MAP} decoder would compute
\begin{align} \label{eq:MAP}
    \hat{\bm{e}} & = \argmax_{\bm{e}\in \mathbb{F}_2^n}{\Pr(\bm{e}|\Tilde{\bm{z}})}
     = \argmax_{\bm{e}\in \mathbb{F}_2^n}{\Pr(\Tilde{\bm{z}}|\bm{e}) \Pr(\bm{e})} \\
    & = \argmax_{\bm{e}\in \mathbb{F}_2^n}{\Pr(\Tilde{\bm{z}}|\bm z(\bm{e})) \Pr(\bm{e})}. \label{eq:MAP2}
\end{align}
The expression in~\eqref{eq:MAP2} can be computed using~\eqref{eq:pe} and~\eqref{eq:pze}.

Note again that evaluating~\eqref{eq:MAP2} requires processing all $2^n$ error vectors.
Let us now see how this complexity can be reduced.
First, $\Pr(\Tilde{\bm{z}}|\bm z(\bm{e}))=\Pr(\Tilde{\bm{z}}|\bm s(\bm{e}))$ has to be evaluated for $2^{n-k}$ different syndrome vectors. This is because all the $2^k$ error vectors in a coset differ by a stabilizer and yield the same corrupted state, thus also the same syndrome.
Second, for a given syndrome vector $\bm s$, the lowest weight error vector $\bm{e^*}(\bm s)$ which is consistent with $\bm s$ maximizes the expression in~\eqref{eq:MAP2}.\footnote{We require $\epsilon<0.5$ and in case there are multiple error vectors with the lowest weight, we pick one of them randomly.} Therefore, before decoding, we can determine a one-to-one mapping between $\bm s$  and $\bm e^*(\bm s)$. For this step, we need to check at least $2^{n-k}$ error patterns, but usually less than $2^n$. This step has to be performed once for a given code prior to decoding. Thus, we can reformulate \ac{MAP} decoding as follows
\begin{align} 
    \hat{\bm{s}}   &  = \argmax_{\bm{s}\in \mathbb{F}_2^{n-k}}{\Pr(\Tilde{\bm{z}}|\bm s) \Pr(\bm{e^*}(\bm s))} .\label{eq:MAPred}
\end{align}
The estimated error vector $\bm{\hat e^*}(\bm {\hat s})$ can be directly obtained from $\hat{\bm{s}}$ through the one-to-one mapping.  According to~\eqref{eq:MAPred}, $2^{n-k}<2^n$ syndrome vectors must be processed during decoding.

\section{Experimental Results}\label{sec:results}
We present experimental results for two families of \ac{CSS} codes whose parity-check matrix structure is as in \eqref{eq:CSS}. In both cases, the submatrices $\bm{H}_X$ and $\bm{H}_Z$ represent two equivalent codes. Therefore, similarly to \cite{MacKay_2004}, we only show simulation results for the code represented by $\bm{H}_X$ over the \ac{BSC} with error probability $\epsilon$. Simulations are performed under both \ac{MAP} and degenerate \ac{MAP} decoding for fixed syndrome error probability $\delta$. We determine the probability of decoding failure $P_e$ versus $\epsilon$ by Monte Carlo simulations. 
{A decoding failure is declared whenever a logical error occurs, i.e., when the decoded error pattern is not in the same coset as the one introduced by the channel.}
Note that the redundant rows of the parity-check matrix may have different weights in our experiments. Therefore, for a fixed $q$ in~\eqref{eq:sj_error}, $\delta$ in~\eqref{eq:avg_delta} will change depending on the code that is considered. For a fair comparison, different codes must be compared for different values of $\delta$.

\subsection{$[[16,2]]$ Product Code}

We consider the $[[16,2]]$ quantum product code \cite{ostrev2022classical}. The $8\times 16$ binary matrix $\bm{H}_X$  in~\eqref{eq:CSS} is given by
\begin{align}
\label{HX_product}
\bm{H}_X=\left[\small{
    \begin{array}{c}
1~1~1~1~0~0~0~0~0~0~0~0~0~0~0~0\\
0~0~0~0~1~1~1~1~0~0~0~0~0~0~0~0\\
0~0~0~0~0~0~0~0~1~1~1~1~0~0~0~0\\
0~0~0~0~0~0~0~0~0~0~0~0~1~1~1~1\\
1~0~0~0~1~0~0~0~1~0~0~0~1~0~0~0\\
0~1~0~0~0~1~0~0~0~1~0~0~0~1~0~0\\
0~0~1~0~0~0~1~0~0~0~1~0~0~0~1~0\\
\hline
0~0~0~1~0~0~0~1~0~0~0~1~0~0~0~1\\
    \end{array}}
    \right] .
\end{align}
Note that $\bm{H}_X$ by construction already contains one redundant weight-$4$ row. Additional redundant rows of weight $6$ are generated by exploiting the code structure. By dividing the rows of matrix $\bm{H}_X$ into two sets, namely the first four rows and the remaining rows, it is possible to generate a total of 16 weight-6 rows. This can be achieved by taking any linear combination of one element from each set. As a result, we obtain a $(24,7)$ syndrome error-correcting code \CD with $d_{\mathrm{min}} = 8$. For completeness, the  submatrix $ \bm A$ of the code's generator matrix $\bm{G}_s=[\bm I| \bm A]$ is
\begin{align}
\label{A_prodcut}
\bm A=
\left[\small{
    \begin{array}{c}
1~1~1~1~0~0~0~0~1~0~0~0~1~0~0~0~1\\
1~0~0~0~1~1~1~1~0~0~0~0~1~0~0~0~1\\
1~0~0~0~1~0~0~0~1~1~1~1~0~0~0~0~1\\
1~0~0~0~1~0~0~0~1~0~0~0~1~1~1~1~0\\
1~1~0~0~1~1~0~0~1~1~0~0~1~1~0~0~1\\
1~0~1~0~1~0~1~0~1~0~1~0~1~0~1~0~1\\
1~0~0~1~1~0~0~1~1~0~0~1~1~0~0~1~1
    \end{array}}
    \right] .
\end{align}

\begin{figure}[t]
    \centering
   \includegraphics[width=0.45\textwidth]{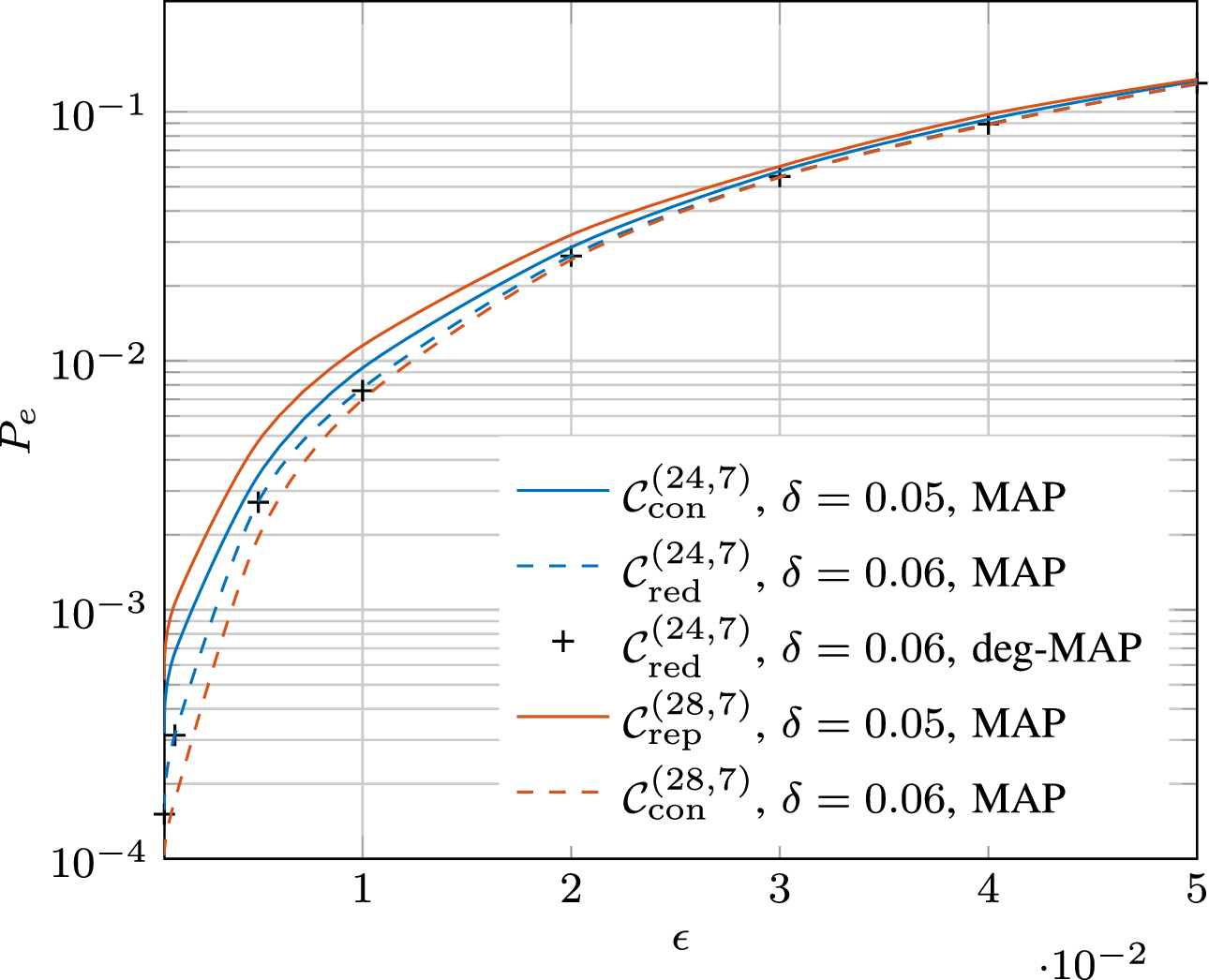}
    \caption{
    Decoding failure rate versus error probability $\epsilon$ for the $[[16,2]]$ product code and $q = 0.013$.}
    \label{fig:product_2}
\end{figure}

We now remove three of the weight-$6$ redundant rows and obtain a $(21,7)$ syndrome error-correcting code \CB with $d_{\mathrm{min}}=6$. Likewise, we consider $\bm{H}_X$ without the last redundant row and repeat the seven syndrome measurements three times. The result is a $(21,7)$ syndrome error-correcting code \CA with $d_{\mathrm{min}}=3$. For a fair comparison between the codes, a higher $\delta$ has to be considered for \CB due to the higher stabilizer weights.
\begin{figure}[t]
    \centering
    \includegraphics[width=0.45\textwidth]{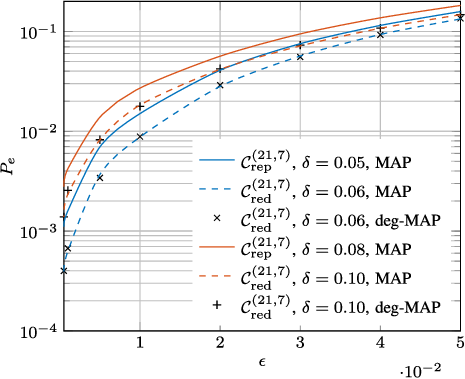}
    \caption{Decoding failure rate versus error probability $\epsilon$ for the $[[16,2]]$ product code with $q = 0.013$ (red) and $q = 0.021$ (blue).}
    \label{fig:product_1}
\end{figure}
Figure \ref{fig:product_1} shows the probability of decoding failure versus $\epsilon$ for different values of $\delta$. While for \CA  we consider $\delta=0.05$ and $\delta=0.08$, for \CB we consider $\delta = 0.0654$ and $\delta =0.1$ due to the additional weight-$6$ rows. Although \CB shows a visible performance gain for both values of $\delta$, \ac{deg-MAP} decoding does not yield performance benefits compared to \ac{MAP} decoding for the current setup. We provide further code design examples of syndrome error-correcting codes. First, we consider $\bm{H}_X$ including the last redundant row, and repeat the eight measurements three times. Formally, the resulting $(24,7)$ syndrome error-correcting code \CC with $d_{\mathrm{min}}=6$ can be described as the serial concatenation of a $(8,7)$ single parity-check code with $d_{\mathrm{min}}=2$ and a $(24,8)$ code with $d_{\mathrm{min}}=3$. The $(24,8)$ code repeats each of the eight information bits three times. Its generator matrix is $\bm I \otimes [1~1~1]$, where $\bm I$ is an $8 \times 8$ identity matrix. Second, we consider $\bm{H}_X$ without the last redundant row and repeat the seven syndrome measurements four times. The result is a $(28,7)$ syndrome error-correcting code \CE with $d_{\mathrm{min}}=4$. Third, we consider \CD and repeat only the first four measurements once. The resulting code is a concatenation of a $(24,7)$ code with $d_{\mathrm{min}}=8$ and a $(28,24)$ code with $d_{\mathrm{min}}=1$. The concatenated code \CF has parameters $(28,7)$ and $d_{\mathrm{min}}=9$. The probability of decoding failure versus $\epsilon$ is shown in Fig.~\ref{fig:product_2}. Again, for a fair comparison, $\delta$ has been adjusted to account for the change in stabilizer weights. Observe that plain repetition of the syndrome measurements leads to noticeable losses in performance. In contrast, using redundant measurements, including also concatenated schemes (with an inner repetition code), leads to an improvement in performance. Again, degenerate \ac{MAP} decoding does not yield visible advantages.

\subsection{$[[18,2]]$ Toric Code}
The second code investigated is the  $[[18,2]]$ toric code \cite{Kitaev97} with
\begin{align}
\label{HX_toric}
\bm{H}_X = \left[\small{
    \begin{array}{c}
1~1~0~0~0~0~0~0~0~1~0~0~0~0~0~1~0~0\\
0~1~1~0~0~0~0~0~0~0~1~0~0~0~0~0~1~0\\
1~0~1~0~0~0~0~0~0~0~0~1~0~0~0~0~0~1\\
0~0~0~1~1~0~0~0~0~1~0~0~1~0~0~0~0~0\\
0~0~0~0~1~1~0~0~0~0~1~0~0~1~0~0~0~0\\
0~0~0~1~0~1~0~0~0~0~0~1~0~0~1~0~0~0\\
0~0~0~0~0~0~1~1~0~0~0~0~1~0~0~1~0~0\\
0~0~0~0~0~0~0~1~1~0~0~0~0~1~0~0~1~0\\
\hline
0~0~0~0~0~0~1~0~1~0~0~0~0~0~1~0~0~1
    \end{array} }
    \right]
\end{align}
$\bm{H}_X$ has only one redundant row. The last row of $\bm{H}_X$ in~\eqref{HX_toric} can be obtained as the sum of all other rows. Overall, we can construct additional $24$ weight-$6$ redundant rows yielding a $(33,8)$ code \CG with $d_{\mathrm{min}}=10$. The  submatrix $ \bm A$ of the code's generator matrix $\bm{G}_s=[\bm I| \bm A]$ is
\begin{align}
\label{A_toric}
\bm A=\left[ \small{
    \begin{array}{c}
1~1~1~1~1~0~0~0~0~0~0~0~0~0~0~1~1~0~0~1~1~1~1~1~1\\
1~1~0~0~0~1~1~1~0~0~0~0~0~0~0~1~0~1~0~1~1~1~1~1~1\\
1~0~1~0~0~1~0~0~1~0~0~0~0~0~0~1~0~0~0~1~0~1~1~1~0\\
1~0~0~1~0~0~0~0~0~1~1~1~0~0~0~0~1~0~1~1~1~1~1~1~1\\
1~0~0~0~0~0~1~0~0~1~0~0~1~1~0~0~0~1~1~1~1~1~1~1~1\\
1~0~0~0~0~0~0~0~1~0~1~0~1~0~0~0~0~0~1~1~0~1~1~0~1\\
1~0~0~0~1~0~0~0~0~0~0~1~0~0~1~0~1~0~0~0~1~1~0~1~1\\
1~0~0~0~0~0~0~1~0~0~0~0~0~1~1~0~0~1~0~0~1~0~1~1~1
    \end{array} }
    \right] .
\end{align}

Next, we show examples of codes with different code parameters. First, removing $9$ weight-$6$ redundant rows, we obtain a $(24,8)$ code \CI with $d_{\mathrm{min}}=6$. Recall that the removal is done such that the minimum distance is kept as large as possible. Similarly, a $(24,8)$ code \CH can be constructed by repeating the first $8$ non-redundant syndrome measurements $3$ times. However, this code has $d_{\mathrm{min}}=3$. Second, removing $6$ rows, we obtain a $(27,8)$ code \CK with $d_{\mathrm{min}}=8$. A $(27,8)$ code \CJ\cmmnt{$(24,8)$ code \CG} can be constructed by repeating all $9$ syndrome measurements. This code can be seen as a serial concatenation of a $(9,8)$ single parity-check code and a code with generator matrix $\bm I \otimes [1~1~1]$, $\bm I$ being a $9\times 9$ identity matrix. Third, removing one weight-$6$ row, we get a $(32,8)$ code \CM with $d_{\mathrm{min}}=9$. A $(32,8)$ code \CL with $d_{\mathrm{min}}=4$ can be obtained by repeating the $8$ non-redundant syndrome measurements $4$ times.
\begin{figure}[t]
    \centering
   \includegraphics[width=0.45\textwidth]{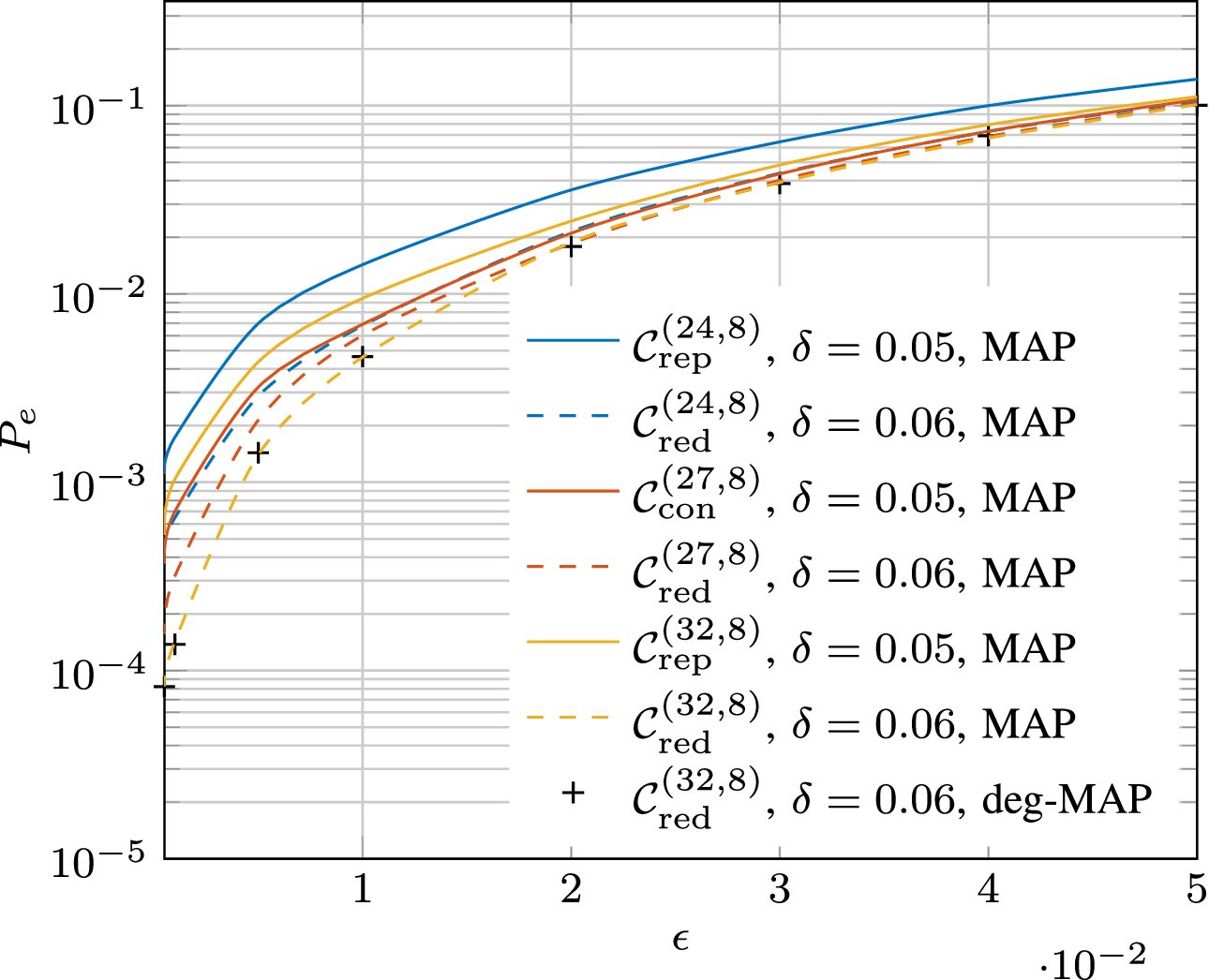}   
    \caption{Decoding failure rate versus error probability $\epsilon$ for the $[[18,2]]$ toric code with $q = 0.013$.}
    \label{fig:toric_1}
\end{figure}
\\The probability of decoding failure versus $\epsilon$ for all codes is shown in Fig.~\ref{fig:toric_1}. Overall, we observe the same trends as for the product code. Degenerate \ac{MAP} decoding does not show visible advantages compared to classical \ac{MAP} decoding. 
 {Also in this case, repeating measurements yields the worst performance.}
 By contrast, the concatenation of a repetition code with other syndrome correction codes {yields good results,
 and it} has the advantage that stabilizer weights can be kept low. The best results for certain code parameters are obtained by choosing appropriate subsets of weight-$6$ stabilizers that define the syndrome error correcting code.

\section{Conclusions}
We studied single-shot decoding of quantum \acf{CSS} codes with faulty syndrome measurements and re-stated the problem as a joint source-channel coding problem. By introducing low-weight redundant rows in the \ac{CSS} code's parity-check matrix, a syndrome error-correcting code is obtained which provides additional resilience against faulty syndrome measurements. By means of code examples, we illustrated that employing a syndrome error-correcting code based on redundant rows outperforms repeated syndrome measurements. Such codes can also be concatenated with an additional repetition code. In our experiments, we considered classical \acf{MAP} decoding, which identifies the most likely Pauli error, and the more complex degenerate \ac{MAP} decoding which subdivides valid error patterns into cosets and identifies the most likely coset. 
In our case, the more  complex degenerate \ac{MAP} decoding turned out to perform similarly to classical \ac{MAP} decoding. 
Experiments with more realistic error models of quantum circuits are left for further work.

\section*{Acknowledgements}
	The authors would like to thank Davide Orsucci for his valuable comments and Gianluigi Liva for helpful discussions.

\bibliographystyle{IEEEtran}
\bibliography{IEEEabrv,references}

\end{document}